\documentclass[journal]{IEEEtran}
\usepackage{cite,graphicx,amsmath,amssymb,psfrag,bm}
\usepackage[usenames,dvipsnames,table,xcdraw]{xcolor}
\usepackage{mathabx}
\usepackage{array}
\usepackage{hyperref}
\usepackage{cancel}
\usepackage{epstopdf}
\usepackage{pstool}
\usepackage{multirow}
\usepackage{float}
\usepackage{color}
\usepackage{soul}
\usepackage{bibentry}
\usepackage{subcaption}

\newcommand{\ve}[1]{\boldsymbol{#1}}

\begin{document}

\title{Efficient GSTC-FDTD Simulation \\ of Dispersive Bianisotropic Metasurface}
\author{Yousef Vahabzadeh, Nima Chamanara
        and~Christophe~Caloz,~\IEEEmembership{Fellow,~IEEE}

\thanks{Y. Vahabzadeh, N. Chamanara and C. Caloz are with the Department
of Electrical Engineering, Polytechnique Montr$\acute{\mathrm{e}}$al, Montr$\acute{\mathrm{e}}$al,
QC, H3T 1J4 Canada (e-mail: yousef.vahabzadeh@polymtl.ca).}
\thanks{Manuscript received MONTH XX, 2016; revised MONTH XX, 2016.}}

\markboth{IEEE Transactions on Antennas and Propagation,~Vol.~X, No.~Y, Month~Z}%
{Shell \MakeLowercase{\textit{et al.}}: Efficient GSTC-FDTD Simulation of Dispersive Bianisotropic Metasurface}

\maketitle


\begin{abstract}
We present a simple and efficient Finite-Difference Time-Domain (FDFD) scheme for simulating dispersive (Lorentz-Debye) bianisotropic metasurfaces. This scheme replaces the conventional FDTD update equations by augmented update equations where the effect of the metasurface, positioned at a virtual node (or node plane) in the Yee grid, is accounted for by judiciously selected auxiliary polarization functions, based on the Generalized Sheet Transition Conditions (GSTCs). This scheme is computationally -- time- and memory-wise -- more efficient and easier to implement than a previously reported scheme for dispersive metasurfaces. It is validated in three illustrative examples.
\end{abstract}


\IEEEpeerreviewmaketitle

\section{Introduction}\label{sec:introduction}
Metasurfaces are engineered subwavelengthly thin $\left(\delta/\lambda_0\ll 1\right)$ materials consisting of a two-dimensional array of scattering particles. In the most general case, they are bianisotropic and nonperiodic~\cite{Kuester_AveragTransCond_2003, Karim_ms_suscp_syn_2015} with effective material parameters that may vary both in space and time~\cite{Nima_simultan_st_spect_2018, Nima_spacetie_proces_ms2016, Caloz_spacetime_ms2016}. They have a myriad of applications, such as for instance, angular momentum conversion~\cite{Capasso_angular_mom_dielec_ms2017}, diffraction-free refraction~\cite{Lavigne_Refr_ms_no_spurious_diffr2018}, harmonic generation~\cite{Karim_2ndorder_nonlinear_ms2017} and antenna radomes~\cite{Eleftheriades_Leaky_ant_ms2017, Esmaeli_Artf_magnet_ms_ant2018}, remote processing~\cite{Karim_remote_proces_2016}, light extraction efficiency enhancement~\cite{Luzhou_Spont_emmision_2016} and solar sails~\cite{Karim_Solar_sail_2017}.

Metasurfaces may be modeled as sheets of zero thickness for simplified design and physical insight~\cite{Alu_wave_transform_grad_ms2016, Maci_Modulated_ms_ant2015,Achouri_birefringent_ms2016, Nima_Exact_polychrom_ms_design2016, Karim_ms_suscp_syn_2015, Xiao_spherical_ms_synthesis2017, Safari_cylinderical_ms2017}. However, no commercial software is available for such structures~\cite{Yousef_Comput_analysis_ms2018}. Therefore, specific numerical techniques have been recently developed for them, both in the frequency domain~\cite{Yousef_Comput_analysis_ms2018, GSTC_FDFD_Yousef_2016, GSTC_FEM_Kumar2017, Nima_SD_IE2017}, namely Finite-Difference Frequency-Domain (FDFD)~\cite{GSTC_FDFD_Yousef_2016, GSTC_FDFD_Yousef_APS2016}, Finite Element Method (FEM)~\cite{GSTC_FEM_Kumar2017} and Spectral Domain Integral Equation (SD-IE)~\cite{Nima_SD_IE2017}, and in the time domain, namely Finite-Difference Time-Domain (FDTD)~\cite{Shulabh_FDTD_broadband_Huygens_ms2017, Shulabh_FDTD_space_time_ms2017, Shulabh_Integr_GSTC_FDTD2017, Achouri_nonlinear_ms2017, Xiao_ms_analys_FDTD2018, GSTC-FDTD_2018_Yousef, Hosseini_PLRCFDTD_GSTC_ms2018}.

FDTD is particularly suited to simulate broadband, time-varying and dispersive structures~\cite{Susan_FDTD2005}. To date, only the FDTD scheme in~\cite{Hosseini_PLRCFDTD_GSTC_ms2018} has included a dispersive treatment of metasurfaces, based on the Piecewise Linear Recursive Convolution (PLRC) technique. However, the formulation in~\cite{Hosseini_PLRCFDTD_GSTC_ms2018} is tedious and computationally inefficient in terms of memory and speed because it involves the inversion of a matrix equation at each time-step.

Here, we present a FDTD scheme that is 1)~exact (no approximation in equation discretization), 2)~efficient in terms of memory and speed, 3)~ applicable to bianisotropic metasurfaces, and 4)~straightforwardly extensible to time-varying dispersive metasurfaces. This method is based on the Auxiliary Differential Equation (ADE) scheme~\cite{Susan_FDTD2005}. In contrast to the conventional ADE for bulk materials, it includes tensorial electric and magnetic polarizations due to bianisotropy. It is therefore more complete but also leads to a more complicated system of equations.

The organization of the paper is as follows. Section~\ref{sec:disp_ms} describes the basic physics of dispersion in materials and provides the related Lorentz, Drude and Debye dispersive models. Section~\ref{sec:ms_synthesis} recalls the metasurface susceptibility GSTC synthesis equations. Section~\ref{sec:formulation} is the core of the paper; it establishes the ADE-dispersive FDTD metasurface analysis. Section~\ref{sec:examples} demonstrates this method via three illustrative examples. Finally, Sec.~\ref{sec:conclusion} draws conclusions.

\section{Dispersive Medium Modeling}\label{sec:disp_ms}
A temporal\footnote{A medium can also be dispersive in terms of the spatial frequency, $\ve{k}$, or spatially dispersive~\cite{Elect_conti_media_Landau_1984}.} frequency dispersive, or temporally dispersive, or, for short, dispersive, medium is a medium whose constitutive parameters depend on the temporal frequency, $\omega$~\cite{Jackson_Classical_Electd2012}. Dispersion is a consequence of causality, which states that any effect must be preceded by a cause~\cite{Nussensveig_caus_disp2012},
 incarnated in the Kramers-Kronig relations~\cite{Jackson_Classical_Electd2012}. The major mechanism leading to dispersion in materials is electronic, atomic, molecular or domain polarizations, which may be macroscopically represented by the electric and magnetic polarization density vectors~\cite{Ishimaru_EM_Radia_Propag_Book1991}.

\begin{figure}[!ht]
\centering
\begin{subfigure}{1\columnwidth}
  \centering
  \includegraphics[width=0.98\columnwidth]{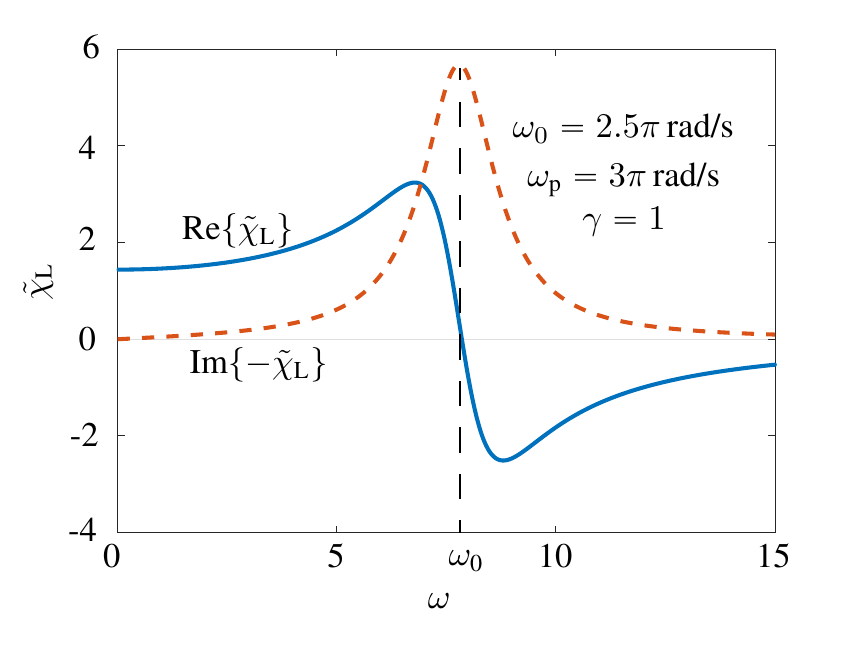}
 \caption{}\label{fig:disp_materials_2}

\begin{subfigure}{1\columnwidth}
  \centering
  \includegraphics[width=1\columnwidth]{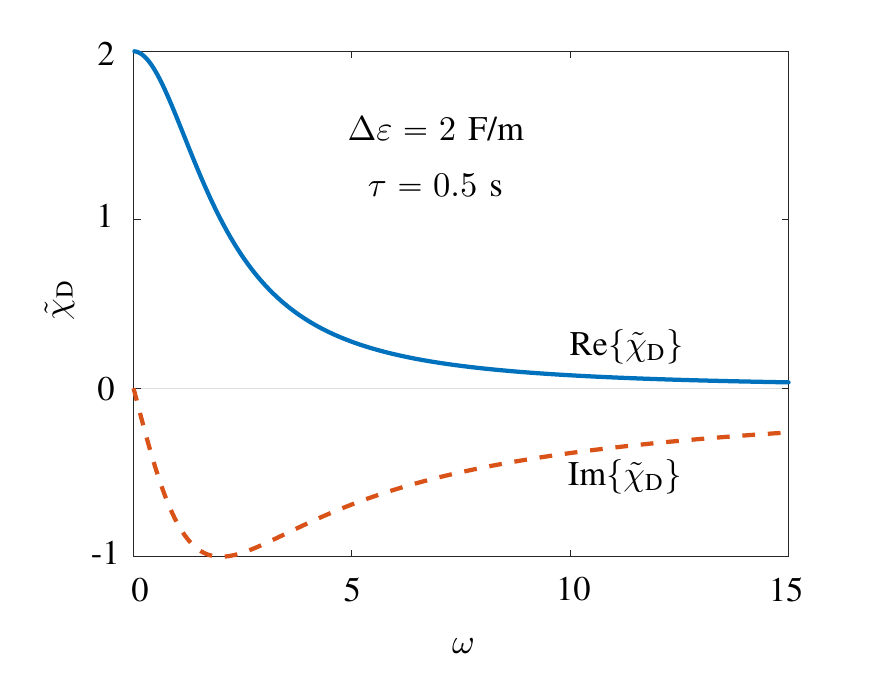}
\caption{}\label{fig:disp_materials_1}
\end{subfigure}
\end{subfigure}
\caption{Complex dispersive susceptibility.~(a) Lorentz model.~(b) Debye model.}
\label{fig:disp_materials}
\end{figure}

Since such polarizations are associated with electron, atom, molecule and domain motions in the medium, the dispersion parameters are found by solving the Newton equation of motion~\cite{Bladel_electromagnetic, Bladel_electromagnetics, Ishimaru_EM_Radia_Propag_Book1991, Jackson_Classical_Electd2012}. This generally leads to the following Lorentz-form dispersion relation in terms of medium susceptibility:
\begin{equation}\label{eq:Lorentz_disp}
  \tilde{\chi}_\textrm{L}(\omega)=\frac{\omega_\textrm{p}^2}{\omega_\textrm{0}^2+2j\omega\gamma-\omega^2},
\end{equation}
where $\omega_\text{p}$ is the plasma frequency, $\omega_0$ is the resonant frequency and $\gamma$ is the damping factor. The real and imaginary parts of $\tilde{\chi}_\textrm{L}(\omega)$ are plotted versus frequency in Fig.~\ref{fig:disp_materials_2}.

In the case of conductors, no resonance occurs since the conduction electrons are not bound, and hence~\eqref{eq:Lorentz_disp} reduces to the Drude dispersion model, $\tilde{\chi}_\textrm{L}(\omega)=\omega_\textrm{p}^2/(2j\omega\gamma-\omega^2)$. In the case of highly lossy materials, such as for instance biological tissues at low frequency, we have $\omega^2\ll\omega\gamma$, and hence~\eqref{eq:Lorentz_disp} reduces to the Debye dispersion
\begin{equation}\label{eq:Debye_disp}
\tilde{\chi}_\textrm{D}(\omega)
=\frac{\Delta \chi}{1+j\omega\tau}
=\chi_\infty+\frac{\chi_\textrm{s}-\chi_\infty}{1+j\omega\tau}
\end{equation}
where $\Delta\chi=(\omega_\text{p}/\omega_0)^2$, $\chi_\textrm{s}$ and $\chi_\infty$ are the static and infinite frequency susceptibilities, respectively, and $\tau=2\gamma/\omega_0$. The real and imaginary parts of $\tilde{\chi}_\textrm{D}(\omega)$ are plotted in Fig.~\ref{fig:disp_materials_1}.

In a metasurface, the scattering particles may include metals, dielectrics or combination of the two. Since these particles are essentially resonators, they also typically exhibit Lorentz or Debye dispersions. Note that while the bulk material susceptibility is unitless the metasurface susceptibility has the unit of meter, as shown in the appendix of~\cite{Xiao_spherical_ms_synthesis2017}.

\section{GSTC Susceptibility Equations}\label{sec:ms_synthesis}
\begin{figure}[!ht]
\centering
\includegraphics[width=0.7\columnwidth]{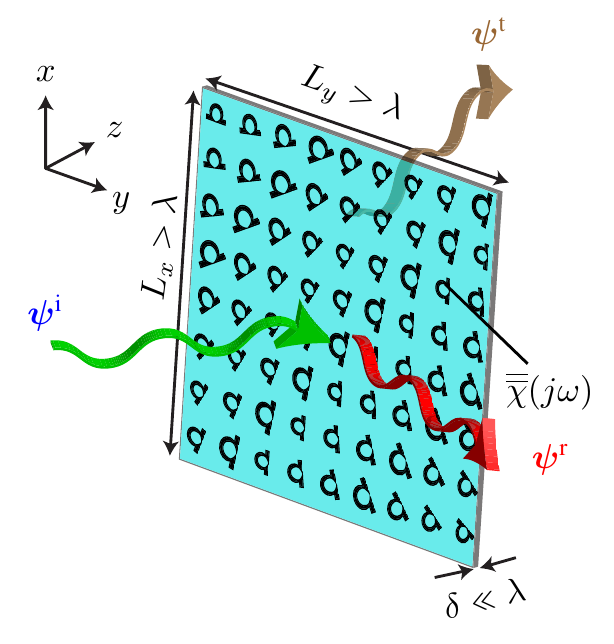}
\caption{Metasurface sheet discontinuity transforming a given incident wave ($\ve{\psi}^\textrm{i}$) into a reflected wave ($\ve{\psi}^\textrm{r}$) and a transmitted wave ($\ve{\psi}^\textrm{t}$).}
\label{fig:ms}
\end{figure}

Figure~\ref{fig:ms} shows a metasurface structure, with key parameters and illustration of field transformation. The corresponding bianisotropic GSTC synthesis equations, assuming only tangential polarizations, are~\cite{Karim_ms_suscp_syn_2015, Kuester_AveragTransCond_2003, Idemen_GSTCBook_2011}
\begin{subequations}\label{eq:freq_GSTC}
  \begin{align}\label{eq:freq_GSTC_1}
  \left(\!\!\!
  \begin{array}{c}
    -\Delta \tilde{H}_y\\
    \Delta \tilde{H}_x\\
  \end{array}
  \!\!\!\right) =&j\omega\varepsilon_0 \left(\!\!
                        \begin{array}{cc}\!\!
                          \tilde{\chi} _{\textrm{ee}}^{xx} & \tilde{\chi}_{\textrm{ee}}^{xy} \\
                          \tilde{\chi}_{\textrm{ee}}^{yx} & \tilde{\chi}_{\textrm{ee}}^{yy} \\
                        \end{array}
                      \!\!\right)
                      \left(\!\!
                        \begin{array}{c}
                          \tilde{E}_{x,\textrm{av}} \\
                          \tilde{E}_{y,\textrm{av}} \\
                        \end{array}
                      \!\!\right)\\\notag
                      &+j\omega\sqrt{\varepsilon_0 \mu_0} \left(\!\!
                                                           \begin{array}{cc}
                                                             \tilde{\chi}_{\textrm{em}}^{xx} & \tilde{\chi}_{\textrm{em}}^{xy} \\
                                                             \tilde{\chi}_{\textrm{em}}^{yx} & \tilde{\chi}_{\textrm{em}}^{yy} \\
                                                           \end{array}
                                                         \!\!\right)
                                                         \left(\!\!
                                                           \begin{array}{c}
                                                             \tilde{H}_{x,\textrm{av}} \\
                                                             \tilde{H}_{y,\textrm{av}} \\
                                                           \end{array}
                                                         \!\!\right),
  \end{align}
  \begin{align}\label{eq:freq_GSTC_2}
  \left(\!\!
  \begin{array}{c}
    \Delta \tilde{E}_y \\
    -\Delta \tilde{E}_x \\
  \end{array}
\!\!\right) =&j\omega\mu_0 \left(\!\!
                        \begin{array}{cc}
                          \tilde{\chi}_{\textrm{mm}}^{xx} & \tilde{\chi}_{\textrm{mm}}^{xy} \\
                          \tilde{\chi}_{\textrm{mm}}^{yx} & \tilde{\chi}_{\textrm{mm}}^{yy} \\
                        \end{array}
                      \!\!\right)
                      \left(\!\!
                        \begin{array}{c}
                          \tilde{H}_{x,\textrm{av}} \\
                          \tilde{H}_{y,\textrm{av}} \\
                        \end{array}
                      \!\!\right)\\\notag
                      &+j\omega\sqrt{\varepsilon_0 \mu_0} \left(\!\!
                                                           \begin{array}{cc}
                                                             \tilde{\chi}_{\textrm{me}}^{xx} & \tilde{\chi}_{\textrm{me}}^{xy} \\
                                                             \tilde{\chi}_{\textrm{me}}^{yx} & \tilde{\chi}_{\textrm{me}}^{yy} \\
                                                           \end{array}
                                                         \!\!\right)
                                                         \left(\!\!
                                                           \begin{array}{c}
                                                             \tilde{E}_{x,\textrm{av}} \\
                                                             \tilde{E}_{y,\textrm{av}} \\
                                                           \end{array}
                                                         \!\!\right),
\end{align}
\end{subequations}
where $\Delta \tilde{\psi}=\tilde{\psi}^\textrm{t}-\left(\tilde{\psi}^\textrm{i}+\tilde{\psi}^\textrm{r}\right)$ and $\Delta \tilde{\psi}=[\tilde{\psi}^\textrm{t}+\left(\tilde{\psi}^\textrm{i}+\tilde{\psi}^\textrm{r}\right)]/2$ with $\tilde{\psi}$ representing any component of the $\tilde{\ve{E}}$ or $\tilde{\ve{H}}$ fields and t, i and r denoting the transmitted, incident and reflected fields, respectively~\footnote{Based on the surface equivalent principle~\cite{Ishimaru_EM_Radia_Propag_Book1991}, any transformation can be represented by equivalent surface currents, leading to only transverse polarization densities. Thus, normal polarization densities lead to redundant solutions, unless one wishes to design a metasurface with different specified fields to different excitations.}.

Equation~\eqref{eq:freq_GSTC} provides the susceptibilities required for a transformation in terms of the specified incident, reflected and transmitted fields. For a single, double or triple transformation, multiple solutions are possible, as discussed in~\cite{Karim_ms_suscp_syn_2015}. Since the solutions are necessarily causal, the metasurface is necessarily dispersive.

We assume that the susceptibilities follow the Lorentz or Debye dispersion models in the bandwidth of interest. Other dispersion models may be handled using expansions in terms of Lorentzian and/or Debye dispersion functions~\cite{VECTFIT_code2002,VECTFIT_method1999}.

\section{Dispersive Metasurface Analysis}\label{sec:formulation}
This section develops a GSTC-FDTD scheme for the simulation of metasurfaces represented by~\eqref{eq:freq_GSTC} with Lorentz~\eqref{eq:Lorentz_disp} or Debye~\eqref{eq:Debye_disp} dispersions. To avoid lengthy equations and tedious developments we consider, without loss of essential generality, a 1D-FDTD problem, i.e. a 0D (point) bianisotropic metasurface with nonzero fields restricted to $(E_y,H_x)\neq0$ and propagation direction $\tilde{\textbf{k}}=k_0\hat{z}$. The extension to the 2D and 3D problems involves a similar procedure, with just more complexity. We assume the general Lorentz dispersion
\begin{equation}\label{eq:ms_DeborLoren}
   \tilde{\chi}_\textrm{ab}(\omega)=\frac{\omega_\textrm{p,ab}^2}{\omega_\textrm{0,ab}^2+2j\omega\gamma_\textrm{ab}-\zeta\omega^2},
\end{equation}
where a can  b can be either e or m~\eqref{eq:freq_GSTC}. The dimensionless coefficient $\zeta$ is used to toggle between Lorentz dispersion $\left(\zeta=1\right)$ and Debye dispersion $\left(\zeta=0\right)$.

\subsection{FDTD Virtual Node}
The conventional 1D-FDTD equations, assuming magnetic and electric fields along the $x$ and $y$ directions, respectively, are~\cite{Susan_FDTD2005}
\begin{subequations}\label{eq:regularFDTD}
   \begin{align}\label{eq:regularFDTD_1}
     &H_x^{n+\frac{1}{2}}\left(i\right)=H_x^{n-\frac{1}{2}}\left(i\right)+\frac{\Delta t}{\mu_0\Delta z}\left[E_y^n\left(i+1\right)-E_y^n\left(i\right)\right],\\\label{eq:regularFDTD_2}
     &E_y^n \left( i\right)=E_y^{n-1}\left( i\right)+\frac{\Delta t}{\varepsilon_0\Delta z}\left[H_x^{n-\frac{1}{2}}\left(i \right) -H_x^{n-\frac{1}{2}}\left( i-1\right)\right],
   \end{align}
\end{subequations}
where $\Delta t=t/n$ and $\Delta z=z/i$ are the FDTD time step and mesh size, respectively.

\begin{figure}[!ht]
\centering
\begin{subfigure}{1\columnwidth}
  \centering
  \includegraphics[width=1\columnwidth]{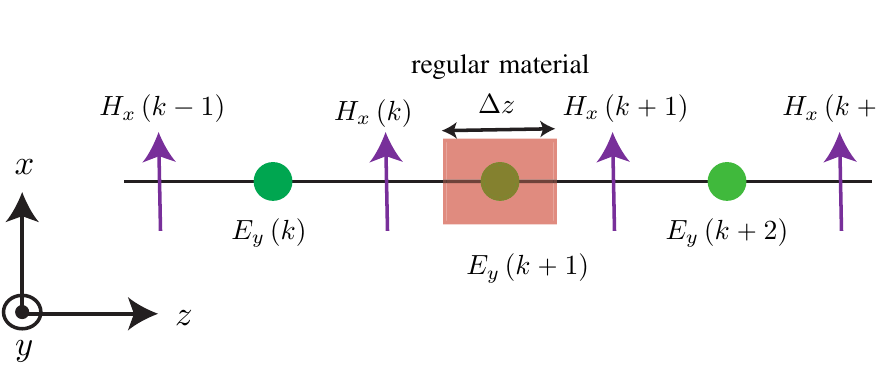}
\caption{}\label{fig:1DFDTD_1}
\end{subfigure}

\begin{subfigure}{1\columnwidth}
  \centering
\includegraphics[width=1\columnwidth]{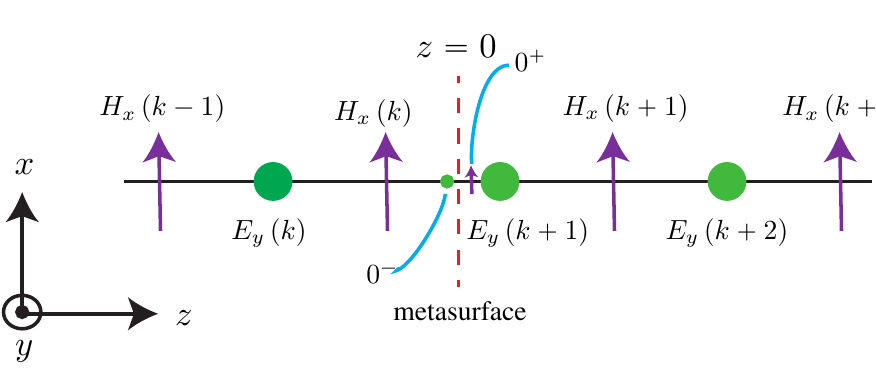}
 \caption{}\label{fig:1DFDTD_2}
\end{subfigure}
\caption{Material positioning in the 1D FDTD grid. (a)~Bulk material, with minimum possible thickness $\Delta z$, positioned at an $E$-field node. (b)~Metasurface, with zero thickness, positioned between adjacent $E$ and $H$-field nodes. The small green circle and purple arrow represent the electric and magnetic virtual nodes placed just before ($z=0^-$) and just after ($z=0^+$) the metasurface, respectively. The metasurface is illuminated from the left in the $z-$direction.}
\label{fig:1DFDTD}
\end{figure}

In the FDTD grid, bulk 3D materials are terminated at either an $E$ or an $H$-field node, and are hence at least one grid cell $\left(\Delta z\right)$ thick, as shown in Fig.~\ref{fig:1DFDTD_1}. In contrast, a metasurface, which is ideally modeled as a zero thickness sheet, can be positioned neither at an $E$ nor at an $H$-field node. For this reason, following~\cite{FDTDGraphene} and~\cite{GSTC-FDTD_2018_Yousef}, we position the metasurface \emph{between} two neighboring cells, as shown in Fig.~\ref{fig:1DFDTD_2}.

Equation~\eqref{eq:regularFDTD} is applicable everywhere in the computational domain, except at the $k$ and $k+1$ metasurface discontinuity nodes, whose update equation involves $E_y(k+1), H_x(k+1)$ and $H_x(k)$. To address this discontinuity issue, we introduce a magnetic \emph{virtual node} (small purple arrow in Fig~\ref{fig:1DFDTD_2}). Then, we incorporate this node into~\eqref{eq:regularFDTD_2}, which yields
\begin{align}\label{eq:updateEy}
  E_y^n \left( k+1\right)=&E_y^{n-1}\left( k+1\right)+\\\notag&\frac{\Delta t}{\varepsilon_0\Delta z}\left[H_x^{n-\frac{1}{2}}\left(k+1 \right) -H_x^{n-\frac{1}{2}}\left(0^+\right)\right].
\end{align}

Similarly, Eq.~\eqref{eq:regularFDTD_1} becomes at the metasurface discontinuity
\begin{equation}\label{eq:updateHx}
  H_x^{n+\frac{1}{2}}\left(k\right)=H_x^{n-\frac{1}{2}}\left(k\right)+\frac{\Delta t}{\mu_0\Delta z}\left[E_y^n\left(0^-\right)-E_y^n\left(k\right)\right].
\end{equation}

Now, $H_x(0^+)$ and $E_y(0^-)$ are computed through the GSTCs~\eqref{eq:freq_GSTC}, which reduce here to
\begin{subequations}\label{eq:simp_GSTC}
    \begin{align}
     \Delta \tilde{H}_x &= j\omega\varepsilon_0 \tilde{\chi}_\textrm{ee}^{yy}\tilde{E}_{y,\textrm{av}}+ jk_0 \tilde{\chi}_\textrm{em}^{yx}\tilde{H}_{x,\textrm{av}},\label{eq:simp_GSTC_1} \\
     \Delta \tilde{E}_y &= j\omega\mu_0 \tilde{\chi}_\textrm{mm}^{xx}\tilde{H}_{x,\textrm{av}}+ jk_0 \tilde{\chi}_\textrm{me}^{xy}\tilde{E}_{y,\textrm{av}},\label{eq:simp_GSTC_2}
   \end{align}
\end{subequations}
with the $\chi$'s given in~\eqref{eq:ms_DeborLoren}.

Since the overall FDTD simulation is performed in the time domain, Eq.~\eqref{eq:simp_GSTC} must be converted into its time-domain counterpart. Such a conversion generally transforms simple products into convolution products, which are often problematic to handle. However, the convolution products may be avoided in particular cases, such as the one relevant here with the Lorentz dispersive function, as will be seen next.


\subsection{Auxiliary Functions}
Due to the staggered nature of the Yee grid, the discretized version of the time-domain equation~\eqref{eq:simp_GSTC} involves mismatch between space and time sampling. The conventional solution in bulk and non-bianisotropic dispersive media is to use the technique of Auxiliary Differential Equations (ADEs)~\cite{Susan_FDTD2005}. Here we extend this technique to bianisotropic metasurfaces. This requires judicious selection of the half-integer and full-integer time steps. Trial and error searching led to the following auxiliary polarization functions\footnote{Note that these functions are not trivial. They must satisfy two essential ADE requirements: 1)~their substitution into~\eqref{eq:simp_GSTC} should lead to a discretizable equation, and 2)~the corresponding ADE should be numerically stable. For example, we numerically found that the auxiliary functions
 \begin{subequations}
  \begin{align*}
  \tilde{P}_\textrm{ee}^{yy} & =\varepsilon_0\tilde{\chi}^{yy}_\textrm{ee}\tilde{E}_{y,\textrm{av}}, \\
  \tilde{P}_\textrm{em}^{yx} & =\frac{\tilde{\chi}^{yx}_\textrm{em}}{c_0} \tilde{H}_{x,\textrm{av}}, \\
  \tilde{M}_\textrm{mm}^{xx} & =\mu_0\tilde{\chi}^{xx}_\textrm{mm}\tilde{H}_{x,\textrm{av}}, \\
  \tilde{M}_\textrm{me}^{xy} & =\frac{\tilde{\chi}^{xy}_\textrm{me}}{c_0} \tilde{E}_{y,\textrm{av}},
\end{align*}
\end{subequations}
used in~\cite{Susan_FDTD2005} for the simulation of bulk dispersive materials, result in unstable update equations.}
\begin{subequations}\label{eq:PM_Lorentz}
  \begin{align}
  \tilde{P}_\textrm{ee}^{yy} & =j\omega\varepsilon_0\tilde{\chi}^{yy}_\textrm{ee}\tilde{E}_{y,\textrm{av}},\label{eq:PM_Lorentz_1} \\
  \tilde{P}_\textrm{em}^{yx} & =jk_0\tilde{\chi}^{yx}_\textrm{em}\tilde{H}_{x,\textrm{av}},\label{eq:PM_Lorentz_2} \\
  \tilde{M}_\textrm{mm}^{xx} & =j\omega\mu_0\tilde{\chi}^{xx}_\textrm{mm}\tilde{H}_{x,\textrm{av}},\label{eq:PM_Lorentz_3} \\
  \tilde{M}_\textrm{me}^{xy} & =jk_0\tilde{\chi}^{xy}_\textrm{me}\tilde{E}_{y,\textrm{av}},\label{eq:PM_Lorentz_4}
\end{align}
\end{subequations}
whose form was inspired -- but modified! -- from the functions involved in the conventional ADE scheme, which uses electric polarization currents as the auxiliary functions~\cite{Susan_FDTD2005}.

Let us verify the validity of the auxiliary functions~\eqref{eq:PM_Lorentz}. Substituting them as an ansatz into~\eqref{eq:simp_GSTC} interestingly yields the coefficient-free relations
\begin{subequations}\label{eq:GSTC_Lorentz}
  \begin{align}
  \Delta \tilde{H}_x &= \tilde{P}_\textrm{ee}^{yy}+\tilde{P}_\textrm{em}^{yx},\label{eq:GSTC_Lorentz_1} \\
  \Delta \tilde{E}_y &= \tilde{M}_\textrm{mm}^{xx}+\tilde{M}_\textrm{me}^{xy}.\label{eq:GSTC_Lorentz_2}
\end{align}
\end{subequations}

Inverse Fourier transforming~\eqref{eq:GSTC_Lorentz_1} and its discretization provides the the time-domain quantity $H_x^{n - \frac{1}{2}}(0^+)$
\begin{align}\label{eq:Hz0p_Lorentz}
  H_x^{n - \frac{1}{2}}\left(0^ +\right) = H_x^{n - \frac{1}{2}}(k) &+\frac{P_\textrm{ee}^{yy,n}+P_\textrm{ee}^{yy,n-1}}{2}+\\\notag &\frac{P_\textrm{em}^{yx,n}+P_\textrm{em}^{yx,n-1}}{2},
\end{align}
whose substitution into~\eqref{eq:updateEy} yields
\begin{align}\label{eq:Ey_update}
E_y^{n}&\left(k +1\right)=E_y^{n-1}\left( k+1\right)
+\frac{\Delta t}{\varepsilon_0\Delta z}\left[H_x^{n-\frac{1}{2}}\left(k+1\right)-H_x^{n-\frac{1}{2}}\left(k\right)\right]
\\\notag&-\frac{\Delta t}{\varepsilon_0\Delta z}\frac{P_\textrm{ee}^{yy,n}+P_\textrm{ee}^{yy,n-1}+P_\textrm{em}^{yx,n}+P_\textrm{em}^{yx,n-1}}{2}.
\end{align}
The first line of this equation is recognized as the conventional FDTD update equation~\eqref{eq:regularFDTD_2}, while the second-line term corresponds the effect of the metasurface discontinuity.

As shown in Appendix~\ref{app:Lorentz_AF_descret}, the auxiliary functions $P_\textrm{ee}^{yy,n}$ and $P_\textrm{em}^{yx,n}$, or ADEs, are obtained from the discretization of~\eqref{eq:PM_Lorentz_1} and~\eqref{eq:PM_Lorentz_2}, respectively, as
\begin{subequations}\label{eq:Pee_n_Lorentz}
\begin{align}\label{eq:Pee_n_Lorentz_1}
  P_\textrm{ee}^{yy,n}&=-\frac{\Delta t^2\omega^2_\textrm{0,ee}-2\zeta}{\Delta t\gamma_\textrm{ee} +\zeta}P_\textrm{ee}^{yy,n-1}-\frac{\zeta-\Delta t\gamma_\textrm{ee}}{\zeta+\Delta t\gamma_\textrm{ee}}P_\textrm{ee}^{yy,n-2}\\\notag&+\frac{\varepsilon_0\Delta t\omega^2_\textrm{p,ee}}{2(\gamma_\textrm{ee}\Delta t+\zeta)}\left[ E_{y,\textrm{av}}^n-E_{y,\textrm{av}}^{n-2}\right],
  \end{align}
  \begin{align}\label{eq:Pee_n_Lorentz_2}
  P_\textrm{em}^{yx,n}&=-\frac{\Delta t^2\omega^2_\textrm{0,em}-2\zeta}{\Delta t\gamma_\textrm{em} +\zeta}P_\textrm{em}^{yx,n-1}-\frac{\zeta-\Delta t\gamma_\textrm{em}}{\zeta+\Delta t\gamma_\textrm{em}}P_\textrm{em}^{yx,n-2}\\\notag&+\frac{\Delta t\omega^2_\textrm{p,em}}{c_0(\gamma_\textrm{em}\Delta t+\zeta)}\left[ H_{x,\textrm{av}}^{n-\frac{1}{2}}-H_{x,\textrm{av}}^{n-\frac{3}{2}}\right].
\end{align}
\end{subequations}

Updating $E_y^n(k+1)$ in~\eqref{eq:Ey_update} requires the knowledge of $P_\textrm{ee}^{yy,n}$, which, from~\eqref{eq:Pee_n_Lorentz_1}, itself depends on $E_y^n(k+1)$ via $E^n_{y,\text{av}}$ according to Fig.~\ref{fig:1DFDTD}. Substituting~\eqref{eq:Pee_n_Lorentz_1} into~\eqref{eq:Ey_update}, and solving for $E_y^n(k+1)$ yields then
\begin{align}\label{eq:Ey_final_update}
E_y^n&(k+1)\left[1+\frac{\Delta t^2\omega^2_\textrm{p,ee}}{8\Delta z(\gamma_\textrm{ee}\Delta t+\zeta)}\right]=E_y^{n-1}\left( k+1\right)
\\\notag&+\frac{\Delta t}{\varepsilon_0\Delta z}\left[H_x^{n-\frac{1}{2}}\left(k+1\right)-H_x^{n-\frac{1}{2}}\left(k\right)\right]\\\notag&-\frac{\Delta t^2\omega^2_\textrm{p,ee}}{4\Delta z(\gamma_\textrm{ee}\Delta t+\zeta)}\left[ \frac{E_{y}^n(k)}{2}+E_{y,\textrm{av}}^{n-2}\right]+c_1P_\textrm{ee}^{yy,n-1}\\\notag &+c_2P_\textrm{ee}^{yy,n-2}-\frac{\Delta t}{\varepsilon_0\Delta z}\frac{P_\textrm{em}^{yx,n}+P_\textrm{em}^{yx,n-1}}{2},
\end{align}
where $c_1=\frac{\Delta t}{2\varepsilon_0\Delta z}\left[-1+\frac{\Delta t^2\omega^2_\textrm{p,ee}-2\zeta}{\gamma_\textrm{ee}\Delta t+\zeta}\right], c_2=\frac{\Delta t}{2\varepsilon_0\Delta z}\frac{-\gamma_\textrm{ee}\Delta t+\zeta}{\gamma_\textrm{ee}\Delta t+\zeta}$, and $P_\textrm{ee}^{yy,n-1}$ is found upon replacing $n$ by $n-1$ in~\eqref{eq:Pee_n_Lorentz_1}.

$E_y(0^-)$ in~\eqref{eq:updateHx} can be handled in an analogous manner using the time-domain version of~\eqref{eq:GSTC_Lorentz_2}, which leads to
\begin{align}\label{eq:Ey0n_Lorentz}
    E_y^n\left(0^-\right) = E_y^n\left(k +1\right) &-\frac{M_\textrm{mm}^{xx,n+\frac{1}{2}}+M_\textrm{mm}^{xx,n-\frac{1}{2}}}{2}\\\notag &-\frac{M_\textrm{me}^{xy,n+\frac{1}{2}}+M_\textrm{me}^{xy,n-\frac{1}{2}}}{2},
\end{align}
whose substitution into~\eqref{eq:updateHx} yields
\begin{align}\label{eq:Hx_update}
  H_x^{n+\frac{1}{2}}&\left(k\right)=H_x^{n-\frac{1}{2}}\left(k\right)+\frac{\Delta t}{\mu_0\Delta z}\left[E_y^n\left(k+1\right)-E_y^n\left(k\right)\right]-\\\notag &\frac{\Delta t}{\mu_0 \Delta z}\frac{M_\textrm{mm}^{xx,n+\frac{1}{2}}+M_\textrm{mm}^{xx,n-\frac{1}{2}}+ M_\textrm{me}^{xy,n+\frac{1}{2}}+M_\textrm{me}^{xy,n-\frac{1}{2}}}{2},
\end{align}
where the first line is the conventional FDTD update equation~\eqref{eq:regularFDTD_1}, while  the second-line term corresponds the effect of the metasurface. Similar to~\eqref{eq:Pee_n_Lorentz}, the auxiliary functions $M_\textrm{mm}^{xx,n+\frac{1}{2}}$ and $M_\textrm{me}^{xy,n+\frac{1}{2}}$ are obtained from discretization of~\eqref{eq:PM_Lorentz_3} and~\eqref{eq:PM_Lorentz_4}, respectively, as
\begin{subequations}\label{eq:Mmm_n_Lorentz}
\begin{align}\label{eq:Mmm_n_Lorentz_1}
    &M_\textrm{mm}^{xx,n+\frac{1}{2}}=-\frac{\Delta t^2\omega^2_\textrm{0,mm}-2\zeta}{\Delta t\gamma_\textrm{mm} +\zeta}M_\textrm{mm}^{xx,n-\frac{1}{2}}-\\\notag&\frac{\zeta-\Delta t\gamma_\textrm{mm}}{\zeta+\Delta t\gamma_\textrm{mm}}M_\textrm{mm}^{xx,n-\frac{3}{2}}+\frac{\mu_0\Delta t\omega^2_\textrm{p,mm}}{2(\gamma_\textrm{mm}\Delta t+\zeta)}\left[ H_{x,\textrm{av}}^{n+\frac{1}{2}}-H_{x,\textrm{av}}^{n-\frac{3}{2}}\right],
\end{align}
  \begin{align}\label{eq:Mmm_n_Lorentz_2}
    &M_\textrm{me}^{xy,n+\frac{1}{2}}=-\frac{\Delta t^2\omega^2_\textrm{0,me}-2\zeta}{\Delta t\gamma_\textrm{me}+\zeta}M_\textrm{me}^{xy,n-\frac{1}{2}}\\\notag&-\frac{-\Delta t\gamma_\textrm{me}+\zeta}{\Delta t\gamma_\textrm{me} +\zeta}M_\textrm{me}^{xy,n-\frac{1}{2}}+\frac{\Delta t\omega^2_\textrm{p,me}}{c_0(\gamma_\textrm{me}\Delta t+\zeta)}\left[ E_{y,\textrm{av}}^{n}-E_{y,\textrm{av}}^{n-1}\right].
  \end{align}
\end{subequations}
Then, updating $H_x^{n+\frac{1}{2}}\left(k\right)$ in~\eqref{eq:Hx_update} requires the knowledge of $M_\textrm{mm}^{xx,n+\frac{1}{2}}$, which, from~\eqref{eq:Mmm_n_Lorentz_1}, itself depends on $H_x^{n+\frac{1}{2}}\left(k\right)$. Substituting~\eqref{eq:Mmm_n_Lorentz_1} into~\eqref{eq:Hx_update}, and then solving for $M_\textrm{mm}^{xx,n+\frac{1}{2}}$ finally yields
\begin{align}\label{eq:Hx_final_update}
  &H_x^{n+\frac{1}{2}}\left(k\right)\left[1+\frac{\Delta t^2\omega^2_\textrm{p,mm}}{8\Delta z(\gamma_\textrm{mm}\Delta t+\zeta)}\right]=H_x^{n-\frac{1}{2}}\left(k\right)\\\notag&+\frac{\Delta t}{\mu_0\Delta z}\left[E_y^n\left(k+1\right)-E_y^n\left(k\right)\right]\\\notag &+c_3M_\textrm{mm}^{xx,n-\frac{1}{2}}+\frac{\Delta t^2\omega^2_\textrm{p,mm}}{4\Delta z(\gamma_\textrm{mm}\Delta t+\zeta)}\left[ \frac{H_{x}^{n+\frac{1}{2}}(k+1)}{2}-H_{x,\textrm{av}}^{n-\frac{3}{2}}\right]\\\notag&+c_4M_\textrm{mm}^{xx,n-\frac{3}{2}}+\frac{\Delta t}{\mu_0 \Delta z}\frac{M_\textrm{me}^{xy,n+\frac{1}{2}}+M_\textrm{me}^{xy,n-\frac{1}{2}}}{2},
\end{align}
where $c_3=\frac{\Delta t}{2\varepsilon_0\Delta z}\left[-1+\frac{\Delta t^2\omega^2_\textrm{p,mm}-2\zeta}{\gamma_\textrm{mm}\Delta t+\zeta}\right], c_4=\frac{\Delta t}{2\varepsilon_0\Delta z}\frac{-\gamma_\textrm{mm}\Delta t+\zeta}{\gamma_\textrm{mm}\Delta t+\zeta}$ and $M_\textrm{mm}^{yy,n-\frac{1}{2}}$ is found upon replacing $n+\frac{1}{2}$ with $n-\frac{1}{2}$ in~\eqref{eq:Mmm_n_Lorentz_1}.
So Eqs.~\eqref{eq:Ey_final_update} and~\eqref{eq:Hx_final_update} are the final update equations taking into account the effect of the metasurface. If the metasurface is not present, $\omega_\textrm{p,ee}=\omega_\textrm{p,mm}=0$, then, these equations reduce to the conventional FDTD equations.

In summary, the dispersive bianisotropic metasurface problem is solved in FDTD using the update equations~\eqref{eq:Ey_final_update} and~\eqref{eq:Hx_final_update}, which reduce to the conventional update equations~\eqref{eq:regularFDTD} away from the metasurface.

\section{Illustrative Simulation Results}\label{sec:examples}
\begin{figure}[!ht]
\centering
\includegraphics[width=1\columnwidth]{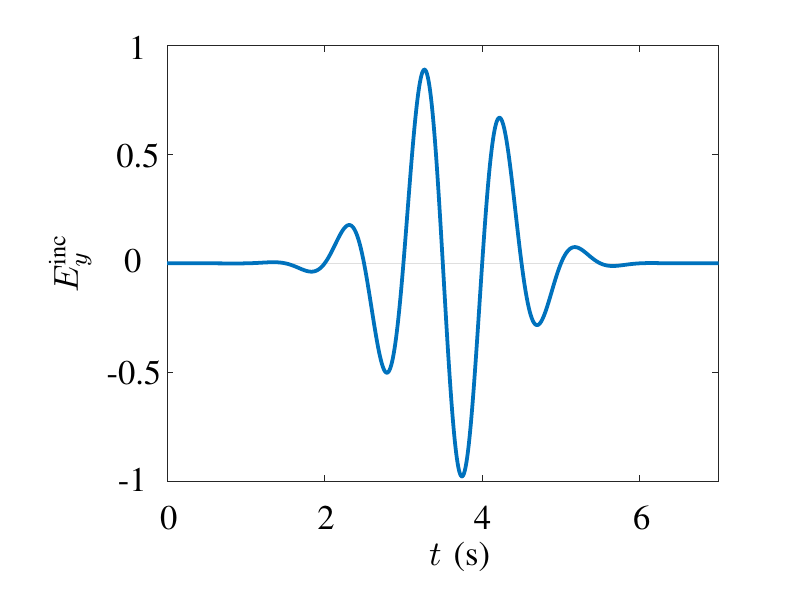}
\caption{Waveform of the incident modulated Gaussian pulse.}
\label{fig:Ey_source}
\end{figure}

All the forthcoming simulations use the normalization $\varepsilon_0=\mu_0=c_0=1$ and $f=1$ Hz with source $E^\textrm{inc}=e^{-(\frac{t-t_0}{\tau})^2}\sin(\omega t)$, plotted in Fig.~\ref{fig:Ey_source}, where $t_0=3.6, \tau=1$ and $\omega=2\pi f$, unless otherwise specified.

\begin{table}[h]
\caption{Summary of the three examples presented in this section. The dimension of the problem is one more than the metasurface dimension.}\label{tb:summary_ex}
  \centering
  \begin{tabular}{|c|c|c|c|}
  \hline
  Nb. & Dispersion & Scattering & dimension and type \\\hline
  1 & Debye & $R,T\neq0$ & 0, bianisotropic \\\hline
  2 & Lorentz & $R=0,T\neq0$ & 0, bianisotropic \\\hline
  3 & Lorentz & $R=0, T=T(y)$ & 1, two anisotropic \\
  \hline
\end{tabular}
\end{table}
Table~\ref{tb:summary_ex} summarizes other parameters of the metasurface. All the results will be compared with the analytic solutions and computed, following the procedures described in~\cite{Karim_ms_suscp_syn_2015} and~\cite{Lavigne_Refr_ms_no_spurious_diffr2018}, as
\begin{subequations}\label{eq:analytic}
  \begin{align}\label{eq:analytic_1}
  S_{11}&=\frac{2jk_0\left(\chi_\textrm{mm}^{xx}-\chi_\textrm{ee}^{yy}+\chi_\textrm{em}^{yx}-\chi_\textrm{me}^{xy}\right)}{2jk_0\left( \chi_\textrm{mm}^{xx}+\chi_\textrm{ee}^{yy}\right)+k_0^2\chi_\textrm{em}^{yx}\chi_\textrm{me}^{xy}+4-k_0^2 \chi_\textrm{mm}^{xx}\chi_\textrm{ee}^{yy}} \\\label{eq:analytic_2}
  S_{12}&=\frac{k_0^2\chi_\textrm{mm}^{xx}\chi_\textrm{ee}^{yy}-\left(2j-k_0\chi_\textrm{em}^{yx}\right) \left(2j-k_0\chi_\textrm{me}^{xy}\right)}{2jk_0\left( \chi_\textrm{mm}^{xx}+\chi_\textrm{ee}^{yy}\right)+k_0^2\chi_\textrm{em}^{yx}\chi_\textrm{me}^{xy}+4-k_0^2 \chi_\textrm{mm}^{xx}\chi_\textrm{ee}^{yy}}
\end{align}
\end{subequations}

\begin{figure}[!ht]
\centering
\includegraphics[width=1\columnwidth]{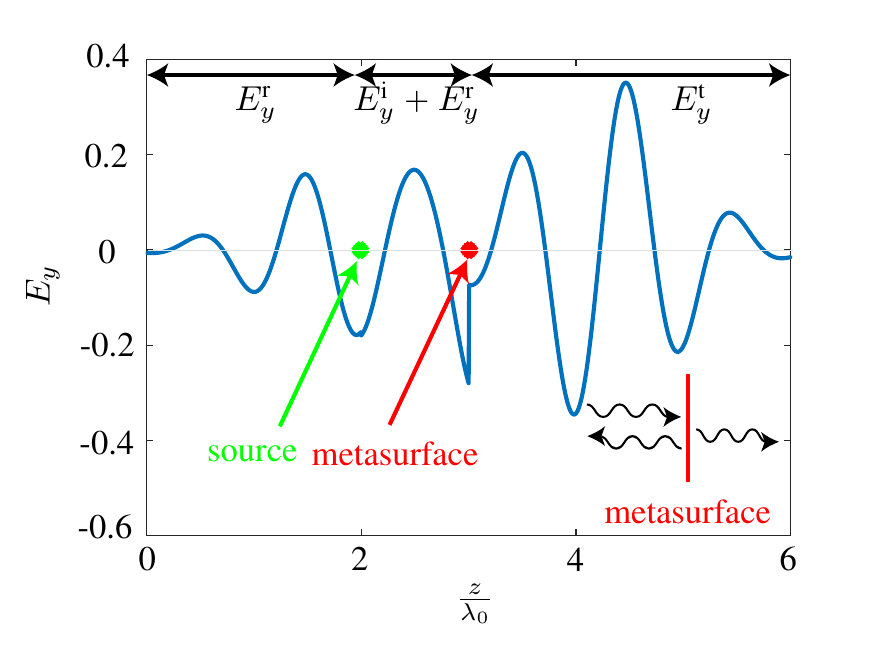}
\caption{Example 1 (Tab.~\ref{tb:summary_ex}): Simulated electric field at time $t=5.8$ s versus space.}\label{fig:Ex1_anisot_Debye_disp_1}
\end{figure}

\begin{figure}[!ht]
\centering
\begin{subfigure}{1\columnwidth}
  \centering
  \includegraphics[width=1\columnwidth]{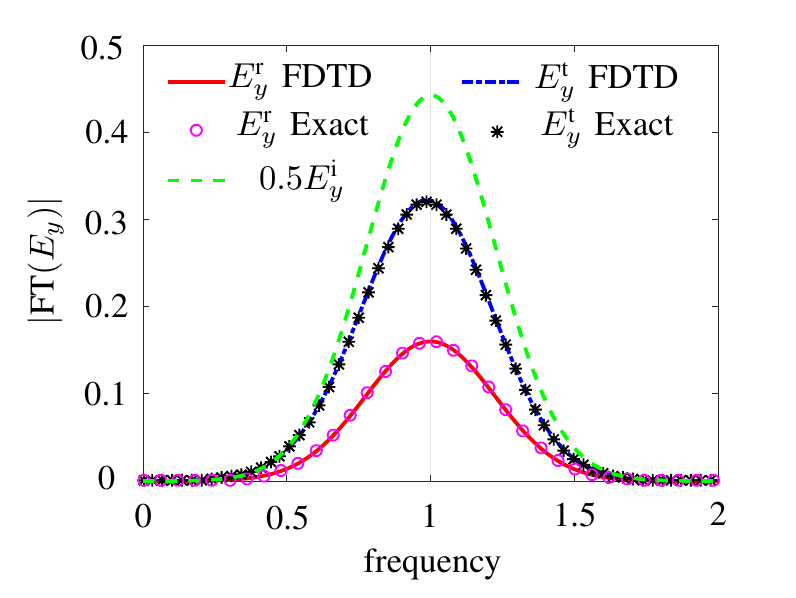}
 \caption{}\label{fig:Ex1_anisot_Debye_disp_2}
\end{subfigure}

\begin{subfigure}{1\columnwidth}
  \centering
  \includegraphics[width=1\columnwidth]{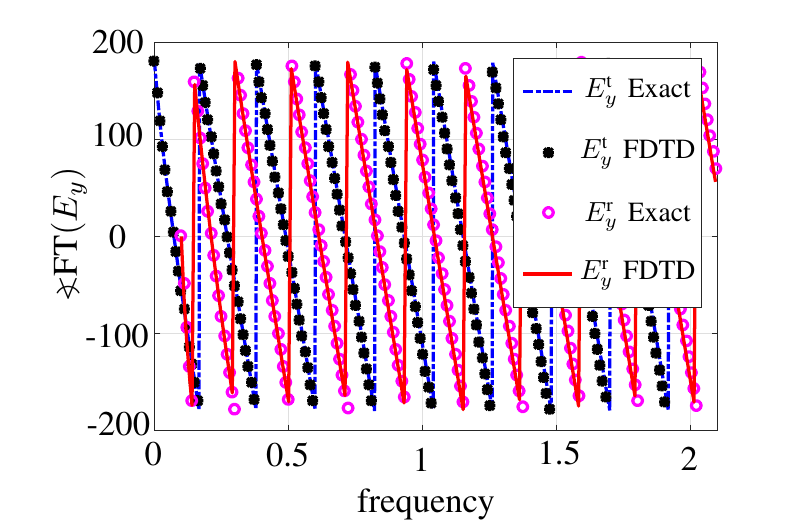}
 \caption{}\label{fig:Ex1_anisot_Debye_disp_3}
\end{subfigure}
\caption{Example 1 (Tab.~\ref{tb:summary_ex}): Fourier transform of the incident and reflected (right before the metasurface) and transmitted (right after the metasurface) electric field in Fig.~\ref{fig:Ex1_anisot_Debye_disp_1} and comparison with the exact result [Eq.~\eqref{eq:analytic}]. (a) Amplitudes. (b) Phases.}
\label{fig:Ex1_anisot_Debye_disp}
\end{figure}

The first example (Tab.~\ref{tb:summary_ex}) involves the Debye dispersive metasurface susceptibilities $\tilde{\chi}_\textrm{me}^{xy}=\frac{2}{1+2j\omega}$ and $\tilde{\chi}_\textrm{mm}^{yy}=\tilde{\chi}_\textrm{ee}^{xx}=\tilde{\chi}_\textrm{em}^{yx}=\frac{2}{1+0.7j\omega}$. The simulation results are shown in Figs.~\ref{fig:Ex1_anisot_Debye_disp_1} and~\ref{fig:Ex1_anisot_Debye_disp}. Figure~\ref{fig:Ex1_anisot_Debye_disp_1} plots the fields in different regions at $t=5.8$~s. According to~\eqref{eq:analytic_1}, the matching condition ($S_{11}=0$) for a bianisotropic metasurface is $\chi_\textrm{mm}^{xx}=\chi_\textrm{ee}^{yy}$ and $\chi_\textrm{em}^{yx}=\chi_\textrm{me}^{xy}$, which is not satisfied in this example. Therefore, the metasurface is mismatched and the reflected field is non-zero ($E_y^\textrm{r}\neq0$). The phase and amplitude of the Fourier transforms of the transmitted and reflected waves, shown in Fig.~\ref{fig:Ex1_anisot_Debye_disp}, are seen to be in agreement with the analytic results obtained from~\eqref{eq:analytic}.

\begin{figure}
\centering
\includegraphics[width=1\columnwidth]{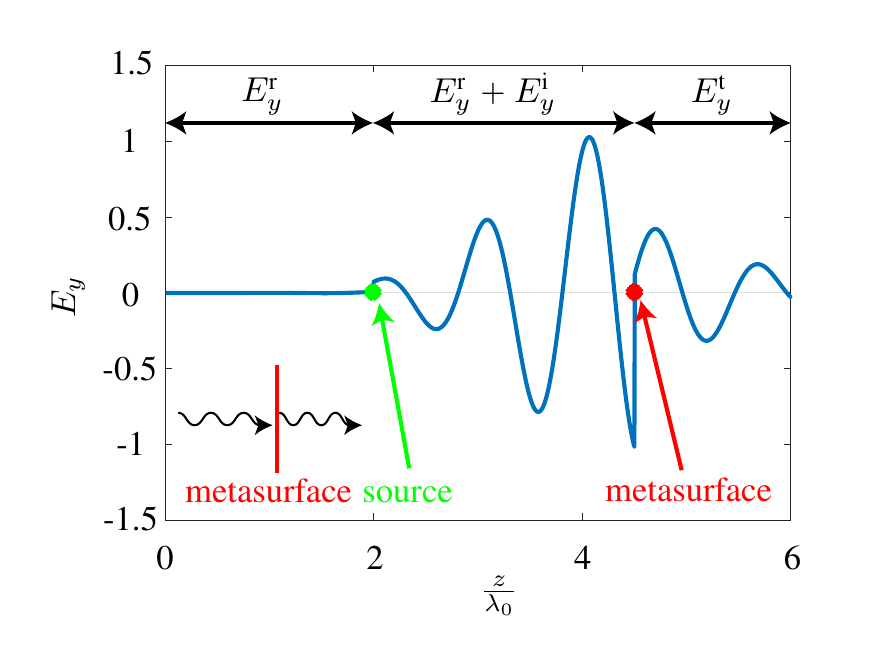}
\caption{Example 2 (Tab.~\ref{tb:summary_ex}): Simulated electric field at time $t=3$ s.}\label{fig:Ex2_anisot_Lorentz_disp_1}
\end{figure}

\begin{figure}
\centering
\begin{subfigure}{1\columnwidth}
  \centering
  \includegraphics[width=1\columnwidth]{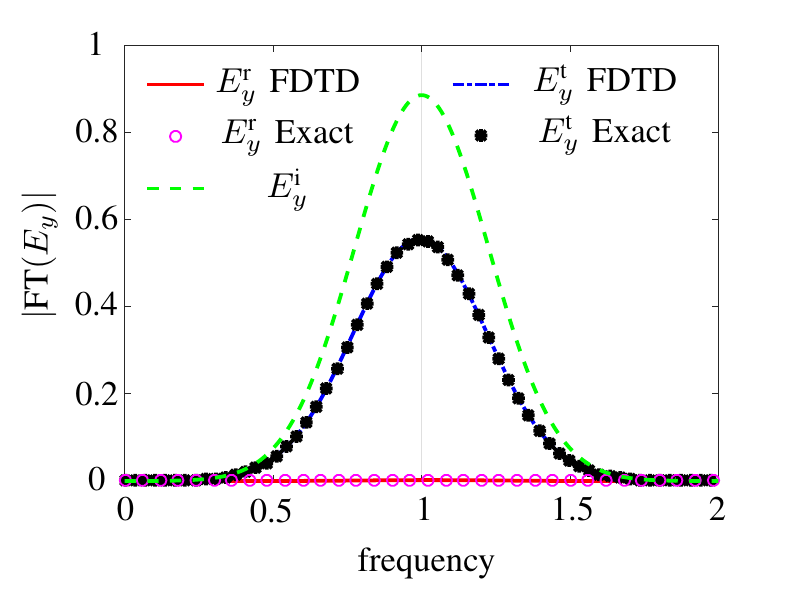}
\caption{}\label{fig:Ex2_anisot_Lorentz_disp_2}
\end{subfigure}

\begin{subfigure}{1\columnwidth}
  \centering
  \includegraphics[width=1\columnwidth]{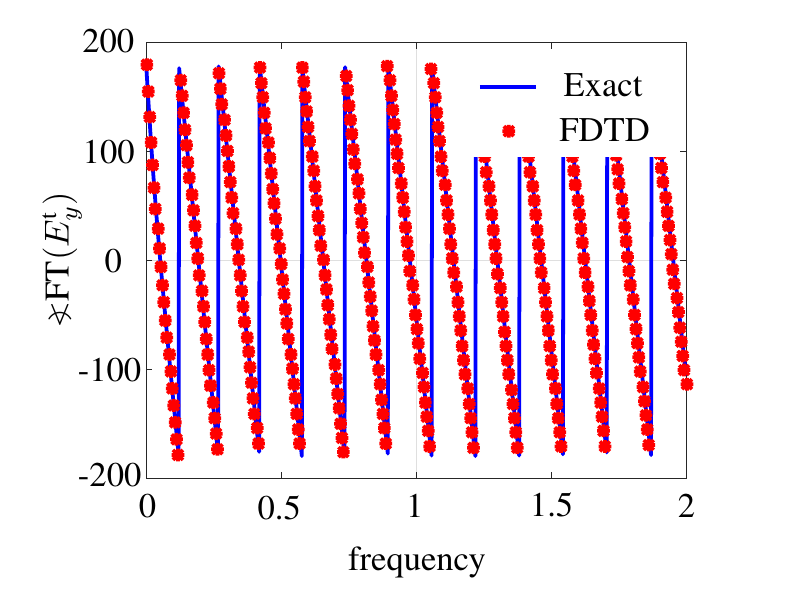}
 \caption{}\label{fig:Ex2_anisot_Lorentz_disp_3}
\end{subfigure}
\caption{Example 2 (Tab.~\ref{tb:summary_ex}): Fourier transform of the incident and reflected (right before the metasurface) and transmitted (right after the metasurface) electric fields in Fig.~\ref{fig:Ex2_anisot_Lorentz_disp_1} and comparison with the exact result, [Eq.~\eqref{eq:analytic}]. (a) Amplitudes. (b) Phases.}
\label{fig:Ex2_anisot_Lorentz_disp}
\end{figure}

The second example (Tab.~\ref{tb:summary_ex}) involves the Lorentz dispersive susceptibilities $\tilde{\chi}_\textrm{ee}^{yy}=\tilde{\chi}_\textrm{mm}^{xx}=\frac{2}{\omega_0^2+2j\omega\gamma-\omega^2}$ and $\tilde{\chi}_\textrm{em}^{yx}=\tilde{\chi}_\textrm{me}^{xy}=\frac{1}{\omega_0^2+2j\omega\gamma-\omega^2}$, where $\omega_0=2\pi20$ and $\gamma=8\omega_0$. Here the matching condition is satisfied and the reflection should therefore be zero. This is verified in Figs.~\ref{fig:Ex2_anisot_Lorentz_disp_1} and~\ref{fig:Ex2_anisot_Lorentz_disp}. The phase and amplitude of the transmitted and reflected fields are again in good agreement with the analytical results.

\begin{figure}
\centering
\includegraphics[width=1\columnwidth]{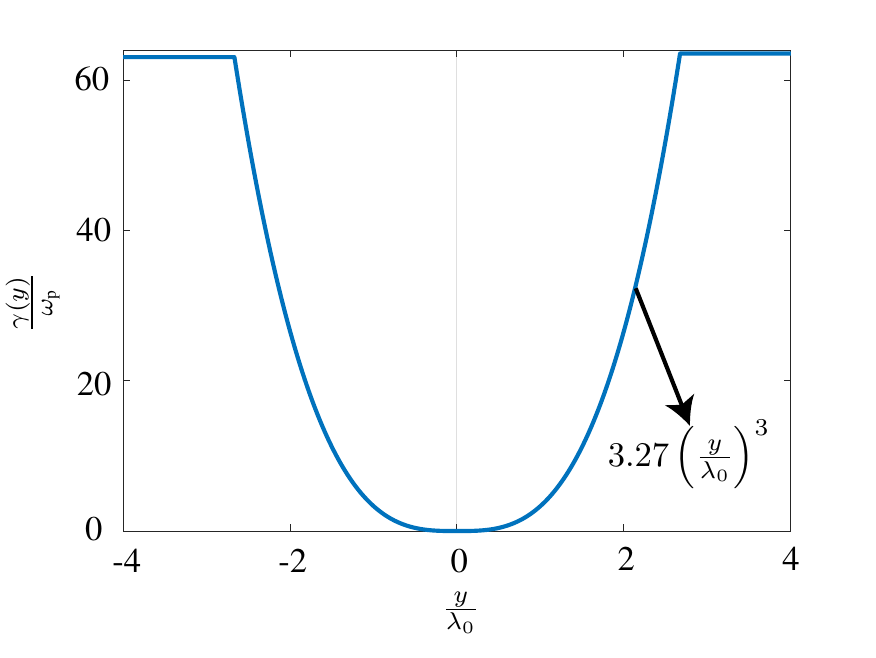}
\caption{Example 3 (Tab.~\ref{tb:summary_ex}): Damping, $\gamma(y)$, profile for full absorption.} \label{fig:Ex3_gamma}
\end{figure}

The third and last example (Tab.~\ref{tb:summary_ex}) involves two parallel space-varying anisotropic metasurfaces excited by a plane-wave incident field ($\tilde{\chi}_\textrm{em}^{yx}=\tilde{\chi}_\textrm{me}^{xy}=0$) with Lorentzian dispersion. The metasurfaces are designed to exhibit the highest transmission at their center and zero transmission at their edges, while being matched with $\tilde{\chi}_\textrm{ee}^{yy}=\tilde{\chi}_\textrm{mm}^{xx}=\frac{\omega_\textrm{p}^2}{\omega_0^2+2j\omega\gamma-\omega^2}$, where $\omega_\textrm{p}=2$ and $\omega_0=2\pi20$. To control the metasurface absorption coefficient, $\gamma$ is varied in space as shown in Fig.~\ref{fig:Ex3_gamma}.

\begin{figure}[!ht]
\centering
\begin{subfigure}{1\columnwidth}
  \centering
  \includegraphics[width=1\columnwidth]{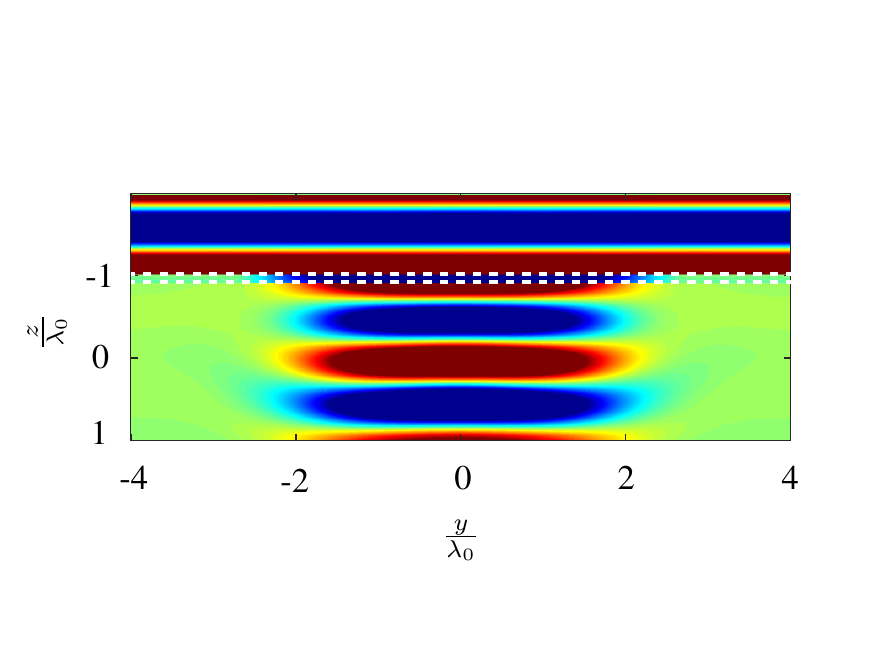}
\caption{}\label{fig:Ex3_anisot_Lorentz_disp_1}
\end{subfigure}

\begin{subfigure}{1\columnwidth}
  \centering
  \includegraphics[width=1\columnwidth]{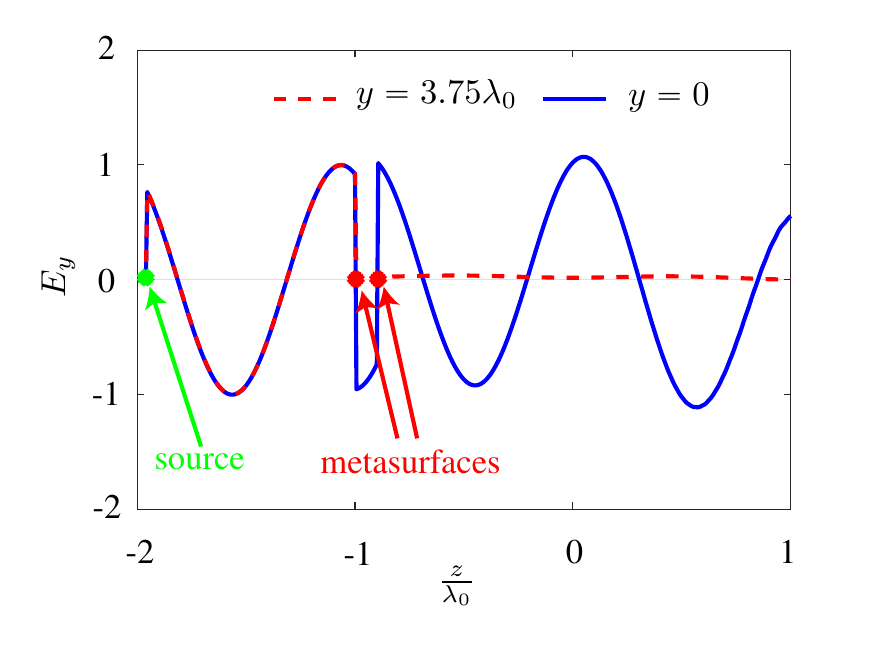}
 \caption{}\label{fig:Ex3_anisot_Lorentz_disp_2}
\end{subfigure}
\caption{Example 3 (Tab.~\ref{tb:summary_ex}): Two-metasurfaces full absorption with illumination in the $+z-$direction. (a)~Field distribution in space, with metasurfaces, in white dashed lines,  located at $z=-\lambda_0$ and $z=-0.9\lambda_0$. (b)~Field distribution in the $z-$direction for $y=0$ and $y=3.75\lambda_0$.}
\label{fig:Ex3_anisot_Lorentz_disp}
\end{figure}

It was numerically found that a single metasurface with Lorentz dispersion cannot absorb all the incident field. Therefore, we stack two metasurfaces and tune their distance for total absorption. This is achieved at $0.1\lambda$, as shown in Fig.~\ref{fig:Ex3_anisot_Lorentz_disp_1}. It can be qualitatively observed that the metasurface exhibits the desired behaviour. This behavior is quantified in Fig.~\ref{fig:Ex3_anisot_Lorentz_disp_2}, where the field distribution at $y=0$ shows almost full transmission with a phase rotation, but zero transmitted field at $y=3.75\lambda_0$, according to specification.

\section{Conclusion}\label{sec:conclusion}

We have presented a simple and efficient Finite-Difference Time-Domain (FDFD) scheme for simulating dispersive -- as well as time-varying and nonlinear -- bianisotropic metasurfaces, using judicious auxiliary polarization functions based on the Generalized Sheet Transition Conditions (GSTCs).

This scheme is a fundamental addition to FDTD. Moreover, it is physically insightful, computationally efficient and easy to implement. For these reasons, its integration into commercial software products, which currently do not effectively allow the simulation of such structures and other emerging complex two-dimensional materials, would be highly beneficial, and may hence become reality in the forthcoming years.

\bibliographystyle{IEEEtran}
\bibliography{TAP_FDTD_Disp}

\appendices
\section{Derivation of~\eqref{eq:Pee_n_Lorentz}}\label{app:Lorentz_AF_descret}
Substituting~\eqref{eq:ms_DeborLoren} for $\tilde{\chi}_\textrm{ee}^{yy}$ and $\tilde{\chi}_\textrm{em}^{yx}$ into~\eqref{eq:PM_Lorentz_1} and~\eqref{eq:PM_Lorentz_2}, respectively, and simplifying, yields
\begin{align}\label{eq:app_Pee_n_Lorentz_S1}
  \left(\omega_\textrm{0,ee}^2+2j\omega\gamma_\textrm{ee}-\zeta\omega^2\right)\tilde{P}_\textrm{ee}^{yy}= \varepsilon_0\omega_\textrm{p,ee}^2j\omega\tilde{E}_{y,\textrm{av}},\\
  \left(\omega_\textrm{0,em}^2+2j\omega\gamma_\textrm{em}-\zeta\omega^2\right)\tilde{P}_\textrm{em}^{yx}= \omega_\textrm{p,em}^2\frac{j\omega}{c_0}\tilde{H}_{x,\textrm{av}}.
\end{align}
The time-domain counterparts of these relations are found by replacing $j\omega$ and $-\omega^2$ by $\frac{d}{dt}$ and $\frac{d^2}{dt^2}$, respectively, which yields
\begin{align}\label{eq:app_Pee_n_Lorentz_S2}
  \left(\omega_\textrm{0,ee}^2+2\gamma_\textrm{ee}\frac{d}{dt}+\zeta\frac{d^2}{dt^2}\right)P_\textrm{ee}^{yy}= \varepsilon_0\omega_\textrm{p,ee}^2\frac{dE_{y,\textrm{av}}}{dt},\\
  \left(\omega_\textrm{0,em}^2+2\gamma_\textrm{em}\frac{d}{dt}+\zeta\frac{d^2}{dt^2}\right)P_\textrm{em}^{yx}= \frac{\omega_\textrm{p,em}^2}{c_0}\frac{dH_{x,\textrm{av}}}{dt}.
\end{align}

Discretization of these equations finally yields
\begin{align}\label{eq:app_Pee_n_Lorentz_S3_1}
  &\omega_\textrm{0,ee}^2P_\textrm{ee}^{yy,n}+2\gamma_\textrm{ee}\frac{P_\textrm{ee}^{yy,n+1}-P_\textrm{ee}^{yy,n-1}}{2\Delta t}+\\\notag&\zeta\frac{P_\textrm{ee}^{yy,n+1}-2P_\textrm{ee}^{yy,n}+P_\textrm{ee}^{yy,n-1}}{{\Delta t}^2}= \varepsilon_0\omega_\textrm{p,ee}^2\frac{E_{y,\textrm{av}}^{n+1}-E_{y,\textrm{av}}^{n-1}}{2\Delta t},
\end{align}
\begin{align}\label{eq:app_Pee_n_Lorentz_S3_2}
    &\omega_\textrm{0,em}^2P_\textrm{em}^{yx,n}+2\gamma_\textrm{em}\frac{P_\textrm{em}^{yx,n+1}-P_\textrm{em}^{yx,n-1}}{2\Delta t}+\\\notag&\zeta\frac{P_\textrm{em}^{yx,n+1}-2P_\textrm{em}^{yx,n}+P_\textrm{em}^{yx,n-1}}{\Delta t^2}= \frac{\omega_\textrm{p,em}^2}{c_0}\frac{H_{x,\textrm{av}}^{n+\frac{1}{2}}-H_{x,\textrm{av}}^{n-\frac{1}{2}}}{\Delta t},
\end{align}
whose resolution for $P_\textrm{ee}^{yy,n}$ and $P_\textrm{em}^{yx,n}$ gives the update equations~\eqref{eq:Pee_n_Lorentz}.
\end{document}